\input harvmac
\def\xa{X^a}
\def\xia{\xi ^a}
\def\htot{ {\cal H}_{tot} }
\def\dsource{ \delta {\cal S}}
\def\hor{ {\cal H}}
\def\rootgamma{\sqrt{\gamma}}
\def\na{n_a}
\def\dpi{\delta \pi}
\def\xhat{ \hat x}
\def\wab{\tilde w _{ab}}

\def\lxi{ {\cal L}_\xi }

\sequentialequations
\lref\jt{ J. Traschen, {\it Constraints on Stress-Energy Perturbations in General
Relativity}, Physical Review D31 (1984) 283. }
\lref\abbott{ L. Abbott and S. Deser {\it Stability Of Gravity With A Cosmological Constant},
Nuclear Physics B {\bf 195}, 76 (1982).}
\lref\kt{  D. Kastor and J. Traschen, {\it Linear Instability of 
Non-vacuum Spacetimes}, Physical Review D47 (1993) 480.}
\lref\wald{D. Sudarsky and R. Wald,{\it Extrema of mass, stationarity, and staticity, and
solutions to the Einstein-Yang-Mills equations} Phys. Rev. D46 (1992) 1453.}
\lref\sorkin{R. Sorkin and M. Varadarajan, {\it Energy extremality in the presence of a
black hole}, Class. Quant. Grav. 13 (1996) 1949.  gr-qc/9410031}
 \lref\bch{J.
Bardeen, B. Carter, and S. Hawking, {\it The Four Laws of Black Hole Mechanics},
 Commun. Math. Phys. 31 (1973) 161.} 
\lref\carter{B.
Carter, in {\it Gravitation and Astrophysics - Cargese 1986}, editors B. Carter and J.
Hartle.}  
\lref\fox{D. Fox, Senior Thesis, Hampshire College (2000).}
\lref\cvetic{M. Cvetic and A. Tseytlin, {\it Nonextreme Black Holes from Nonextreme 
Intersecting M-branes}, Nucl. Phys. B478 (1996) 181, hep-th/9606033.}
\lref\thank{JT would like to thank Robert Wald for pointing this out to me while I was giving
a seminar at University of Chicago, 10/92.}
\lref\arms{J. Arms, J. Math. Phys 18, 830 (1977); J. Math. Phys. 20, 443 (1979)}
\lref\amm{J. Arms, J. Marsden and V. Moncrief, {\it The Structure of the Space of Solutions
of Einstein's Equations II}, Annals of Physics 144 (1982) 81.}
\lref\mon{V. Moncrief, Phys. Rev. D18 (1978) 983.}
\lref\mtw{Misner, Thorne, and Wheeler, {\it Gravitation}, W.H. Freeman and Co. (1973).}
\lref\waldbook{R. Wald, {\it General Relativity}, University of Chicago Press, 1984.}
\lref\tod{P. Tod, {\it The Integral Constraint Vectors of Traschen and Three-Surface 
Twistors}, Gen. Rel. Grav. 20 (1988) 1297. }
\lref\teit{C. Teitleboim, {\it Surface Integrals as Symmetry Generators in Supergravity
theories}, Physics Letters 69B, 240 (1977)}
\lref\horowitz{
G.~T.~Horowitz and D.~L.~Welch,
``Duality invariance of the Hawking temperature and entropy,''
Phys.\ Rev.\ D {\bf 49}, 590 (1994)
[arXiv:hep-th/9308077].
}
\lref\taylor{W. Taylor, {\it D2-Branes in B Fields}, JHEP 0007 (2000) 039, 
hep-th/0004141.}
\lref\town{P. Townsend and M. Zamaklar, {\it The First Law of Black Brane Mechanics},
Classical and Quantum Gravity  {\bf 18}, 5269 (2001), hep-th/0107228}
\lref\regge{ T.Regge and C.Teitelboim,
``Role Of Surface Integrals In The Hamiltonian Formulation Of General Relativity,''
Annals Phys.  {\bf 88}, 286 (1974).}
\lref\teitelboim{ C.Teitelboim,
 "Surface Integrals as Symmetry Generators in Supergravity Theory",
 Physics Letters\ {\bf 69B}, 240 (1977).}
\lref\jtspinor{J. Traschen, ``A Positivity Theorem for Gravitational Tension in Brane Spacetimes",
hep-th/0308173.}

\Title{}{Tension Perturbations of Black Brane Spacetimes}
\centerline{Jennie Traschen \foot{traschen@physics.umass.edu} and
Daniel Fox\foot{dosmanos@math.duke.edu} }
\bigskip\centerline{\sl ${}^1$ Department of Physics, University of Massachusetts, Amherst, MA
01003-4525} 
\medskip\centerline{\sl ${}^2$Hampshire College and Department of Mathematics, Duke University}
\bigskip
\centerline{\bf Abstract}
\bigskip\medskip
We consider black-brane spacetimes that have at least one
 spatial translation Killing field that is tangent to the
brane. A new parameter, the tension of a spacetime, is defined. The tension parameter
is associated with spatial translations in  much the same way that the ADM mass 
is associated with the time translation Killing field. In this work, we
 explore the implications of the spatial translation symmetry for small perturbations around a
background black brane. For static charged black branes
we derive a law which relates  the tension perturbation to the surface gravity times the
change in the the horizon area, plus terms that
involve variations in the charges and currents. We find that as a black brane evaporates 
the tension decreases.  We also give a simple derivation of a first law for black brane spacetimes.
These constructions hold when the background  stress-energy is governed by a Hamiltonian, and
the results include arbitrary perturbative stress-energy sources.

\Date{} 


\newsec{Introduction and Summary of Results}

In this paper, we introduce a new gravitational charge, the spacetime tension.
 The idea of the tension
 of a spacetime is simple. A particle-type object, like a billiard ball, has a rest mass.
 When the mass becomes large and the self-gravity of the object is important, as for a star, 
there are gravitational contributions to the total mass of the system. The mass $M$ of an
asymptotically
 flat spacetime is defined as a boundary integral of the long range gravitational field, and can
 be constructed as a conserved charge associated with the asymptotic time translation Killing
 vector. Now, consider an elastic band, rather than a ball.
 This has  tension as well as mass. Suppose that the elastic band is either infinite or wrapped around
an $S^1$ factor of the spacetime. If the band becomes
self-gravitating, then there will be a gravitational contribution to the tension.
In analogy with mass, the tension $\mu$ of a spacetime is the charge associated with an
asymptotic spatial translation
  Killing vector which is tangent to the band.\foot{ After the first version of this
paper appeared, work of Townsend and Zamaklar 
\town\ appeared which also defined the spacetime `fvztension,
using the covariant techniques of \abbott\ . It is straightforward to check that the two
expressions for the definition of tension agree.
 Recently, one of us has used spinor techniques to prove a
positivity result about gravitational tension \jtspinor\ }.

 The tension of a spacetime arises as an extension of the usual ADM gravitational
charges that are derived from the variation of the gravitational Hamiltonian. Let ${\cal H} _t$ be the
Hamiltonian that generates flow along a timelike vector field . 
The lapse and shift appear as Lagrange multipliers in ${\cal H} _t$.
 Regge and Teitelboim \regge  ,\teitelboim\   computed the variation of  ${\cal H}_t$  
for asymptotically flat spacetimes.  
Evaluated on solutions, it is given by the sum of boundary integrals at spacelike infinity, multiplied
by the Lagrange multipliers. For asymptotically  flat spacetimes, the
asymptotic symmetry group is the Poincare group. They demonstrated that if  
 the Lagrange multipliers are chosen to be asymptotic to the generators of the Poincare group,
then the variation of ${\cal H}_t $ is
the sum of the variation of the gravitational four-momentum and angular momentum,
each multiplied by the corresponding generator.
For example, the variation in the ADM mass
$\delta M$ is the coefficient of an asymptotic time translation.

If we now consider a spacetime 
that is asymptotically flat
cross an $S^1$, then the asymptotic symmetry group has an additional spatial translation 
 Killing vector ${\partial\over \partial x}$. Therefore in the Regge-Teitelboim construction
the variation of  ${\cal H} _t$  contains
   an additional linear momentum term $\delta P^x$. The key step in defining the spacetime
tension is to note that 
because of the covariance  of the theory, one can just as well write down a Hamiltonian
funtion ${\cal H} _x$ which generates flow along a spacelike vector field, asymptotic to
${\partial \over \partial x}$.
The variation in the gravitational tension  $\delta\mu$ is then defined  
 by computing the variation in ${\cal H} _x$ when the Lagrange multiplier is
asymptotic to   ${\partial\over \partial x}$. This construction will be given in
Section 2 below.

 We will also be interested in properties of perturbations to
 the gravitational mass and tension of $p$-brane spacetimes.
When a spacetime has a symmetry, then often one can prove useful relations that hold for
perturbations around the spacetime. The most famous example of this is the first law of black
hole mechanics \bch\ , 
which holds if the background spacetime has a stationary Killing field  $\xi^a$.
In addition to $\xi ^a$, a black p-brane spacetime may have one or more spatial translation Killing
fields $\xa$ tangent to the brane.  It is natural to ask whether there is another
law of black brane mechanics
that relates  variations  in the tension $\delta \mu$, to variations in the 
geometrical properties of the horizon.

In order to address this,
we start in Section 2 by deriving a general integral identity, or constraint, which holds on solutions to
the linearized Einstein equation coupled to matter fields. The derivation assumes that
the background stress-energy comes from a matter Hamiltonian. Perturbative stress-energy
sources are included, and these need not have  a Hamiltonian description. For definiteness, when
writing out explicit formulae, we assume that the background stress-energy comes from a
$(p+1)$-form abelian gauge field, which generally arise in supergravity theories and couple
to$p$-branes.
This general construction is then used to prove three results concerning perturbations of
black $p$-branes. 

First, to illustrate the techniques, in Section 3
we give a simple proof of the first law as it applies to black branes.
For example, let the background
spacetime be a static black 2-brane which is electrically charged with respect to the 3-form
gauge potential $A_{abc}$. Let the directions tangent to the 2-brane be compact.
 Then using the static Killing field $\xi^a$,
our general construction  gives the first law

\eqn\first{\delta M = {\kappa\over 8 \pi}\delta A +\Omega \delta {\cal J}
 -{3\over 2 \pi }A_{tab}  \delta Q^{tab}
+\int_V ( \xia  \delta T ^{(s)}_{\hat{t} a} +4! A_{tcd}\delta j^{tcd} ).}
 Here   $\kappa$ is the surface
gravity, $A$ is the horizon area, $\Omega$ is the angular velocity of the horizon, and
${\cal J}$ is the angular momentum of the spacetime. 
 $\delta Q^{eab} = 4!\int _{\partial V} Nda _c \delta F^{cabe}$ is
the variation of the electric charge, where $F=dA$ .
 The volume integral gives the
contribution of perturbative sources  $\delta T^{(s)a}_{\ \ \
\  b}$ and charged currents $\delta j^{abc}$.

Evaluating \first\ in the case that the background spacetime has no horizons and no gauge fields
gives
\eqn\moffflat{\delta M= \int _V \xia \delta T ^{(s)}_{\hat{t}  a}, }
That is, $\delta M$ is equal to integral of the local energy density, which agrees with our
Newtonian intuition about mass.

 Second, in Section 4 we prove a
statement analogous to equation \first\  for  variations
in the tension $\delta\mu$ of black branes. 
$\delta \mu$ is first defined by a boundary integral
which depends on $\xa$, the background
metric, and the metric perturbation. The functional dependence of $\delta \mu$ on  
 $X^a$ is similiar to the dependence of $\delta M$  on $\xi^a$. 
The main result of this paper is to derive a relation for
$\delta\mu$ which is analogous to equation \first\  .
Consider again the example of a static charged, black 2-brane, and that the brane wraps a 2-torus.
 Assume that the spacetime has a
spatial translation Killing vector $X^a={\partial\over \partial x} $, which is tangent to an $S^1$
wrapped by the brane. Then using $X^a$ in our general construction gives the relation 
\eqn\introlaw{\delta \mu  =- { A\over 8 \pi L_H }\delta \kappa
 -{3\over 2 \pi} A_{xab}  \delta Q^{xab}
+ \int _{V_x} [- \xa \delta T ^{(s)\hat x}_{\ \ \ \ a} 
 +{3\over 2 \pi} A_{bcx}\delta J^{bcx } ] }
 Here $L_H$ is the length of the $S^!$, to which $X^a$ is tangent, at the horizon.
 $V_x$ is a subsurface of an asymptotically flat slice which has $x=constant$.  
Using a Komar-Smarr relation, this can be rewritten as
\eqn\introlawtwo{ \delta \mu  = { \kappa\over 8 \pi (n-2) L_H }\delta A + gauge \  field \  and\ 
source\  terms.}
 
As for $\delta M$, we can gain physical understanding of the quantity $\delta \mu$ by specializing
\introlaw\  to 
 the case of  perturbations off a spacetime with no horizons and no gauge fields. Then

\eqn\poffflat{ \delta\mu = -\int _V  \xa \delta T ^{(s)\hat x}_{\ \ \ \ a}  ,}
$i.e.$, $\delta\mu$ is minus the integral of the $x$-component of the pressure of
the stress-energy.
This is part of the justification of calling the boundary integral the tension.
Note that the momentum associated with the Killing field $X^a$ is a different
physical quantity; in this same example,  $\delta P^x =
 \int _{\Sigma} \xa \delta T ^{(s)}_{\hat{t}  a}  $

 We use the term ``tension" rather than ``pressure"
because test branes whose dynamics are governed by the area action have positive tension. 
Note that the tension is associated with a Killing field tangent to
the brane, and is not a radial pressure as in a star. If the
two-brane has a second translational Killing field $Y^a$, then there is an analogous statement
for variations in the tension associated with the y-direction.

 Third, we compute the variation in the tension $\delta\mu$ when a charged test
brane moves in a background magnetic field. 
Then $\delta\mu$ has contributions from the perturbation to the enclosed current, and from
the variation of the magnetic field on the boundary, as well as the perturbed stress-energy.
The resulting statement looks like a gravitational-Ampere's Law.
This is worked out in Section (4.2) .

We close with some remarks about the relation of this work to some other work which has
used Hamiltonian techniques. References \arms\ and \amm\ studied the linearization instability of
Einstein-Maxwell and Einstein-Yang-Mills on  spacetimes with $compact$ spatial sections. They
 prove that linearization instability occurs when the spacetime has a Killing field.
If the space is non-compact, there is no linearization instability. Instead, their approach,
which we employ here, yields an identity on solutions to the linearized equations. The identity
takes the form of a guass's law constraint, with a boundary term. When the Killing field
of the spacetime is time translation, then the constraint is the first law (see Section 3).

 Reference \wald\  studies Einstein-Yang-Mills theory on spacetimes with non-compact
spatial sections. They use a
Hamiltonian $h$ for Einstein-Yang Mills theory which includes an appropriate boundary term at infinity.
Evaluated on solutions only the boundary term of $h$ is nonzero, and this is interpeted as the 
energy of the spacetime. The perturbations of $h$ are computed and this yields the First
Law. By contrast,
in this paper as in our previous work \jt\kt\  , we start with the field equations. 
The relations \first\ and \introlaw\ are  identities, or constraints,
on solutions to the linearized equations. The challenge  is
to understand the geometrical meaning of the boundary terms. We argue
that a particular boundary integral can be identified as the tension of the spacetime.

The definitions and properties of the gravitational charges in asymptotically flat spacetimes
have been extensively studied.
A nice feature of our approach is that one is not limited to asymptotically flat boundary conditions.
The constraint(s) must be true for perturbations about  any background spacetime which has  
Killing vector(s), such as Reissner-Nordstrom-deSitter or Ernst spacetimes. For example,
consider a black hole in deSitter. Let
 $V$ be a spacelike slice  bounded by the black hole and deSitter horizons. In this region there
is a static Killing field $\xi^a$. 
Using $xi ^a$ in our general construction gives  
the identity  $\kappa _{ds}\delta A_{ds} =\kappa_{bh}\delta A_{bh}$, where $\kappa_{bh} ,\kappa_{ds}$
are the surface gravities of the black hole and deSitter horizons, and $\delta A_{bh},\delta A_{ds}$
the change in the areas of the horizons. The interesting thing is that
the boundary term which gives $\delta M$ on an asymptotically flat boundary becomes
 $\kappa _{ds}\delta A_{ds}$ on the deSitter horizon.
 
This paper is organized as follows. In section 2 we derive  general constraint  relation
on perturbations for a foliation of the spacetime by either spacelike or timelike
surfaces. This relation is applied to prove the first law in section 3, and
then to derive the constraint 
on $\delta\mu$ in section 4.  We show that the $\delta\mu$ constraint simplifies to equation \introlaw
when the background and the perturbations are static.
Some geometrical properties of the horizon, related to the existence of
 a spatial translation Killing field, are derived. Section 5 contains concluding remarks. 
Appendix A contains some details of the Hamiltonian decomposition of the three-form gauge field
Lagrangian. In Appendix B we show that a $p$-form gauge potential is constant on a bifurcate
Killing horizon.

\newsec{Integral Constraints on Perturbations}

Since the idea of the derivation is simple, while the calculations are detailed,
 we first outline the idea. The
Hamiltonian techniques in \jt\kt\wald\  and the covariant techniques in \sorkin
\bch\carter\ are made use of. Consider
the Einstein Lagrangian $R$ coupled to matter fields which are described by a
Lagrangian, for example $L_M = -F^2$. Let the spacetime be foliated
by spacelike slices, with timelike normal vector field
$n=-Ndt$. One can then construct the Hamiltonian which generates
flow along the time direction $\xia ={\partial \over \partial t}$. The Hamiltonian $\htot$ for the
coupled Einstein-Maxwell system  is pure constraint, namely the sum of the Einstein constraints
and Gauss's law, multiplied by lagrange multipliers. On solutions 
$\htot =0$. Perturbative solutions linearized about a background
satisfy $\delta \htot =0$, where the variation is with respect to all the
dynamical fields $``p"$ and $``q"$. The latter  equation can be rewritten in terms of
the adjoint operator $\delta\htot ^{\dagger}$ and a total derivative. That is, one can integrate
the equation over a volume $V$ contained in a spatial slice,
 and integrate by parts. This yields a volume integral of the adjoint,
 plus an integral over the boundary $\partial V$. But the adjoint operator
generates the Hamiltonian flow, and so Hamilton's equations imply that  on solutions
the volume integral is simply
the lie derivative along the direction $\xia$, of the $p's$ and $q's$. If 
$\xia$ is a Killing vector for the background then the lie derivatives vanish and
the result is greatly simplified.
The resulting statement is an identity, or constraint, which must hold on
solutions to the linearized equations.  If the background spacetime is a
 black brane, this construction
gives the first law \thank . To include perturbative sources $\dsource$,
one simply starts with the linearized equation  $\delta \htot=\dsource$.

Next consider foliating the spacetime with slices that have a
{\it spacelike} normal vector field.
This defines a Hamiltonian flow along a spacelike field $\xa$. The
construction of $\htot  ,\delta\htot$ and $\delta\htot ^{\dagger}$ 
is much the same as for the decompositon based on a timelike normal.
 One just has to be careful about various minus signs. 
Again the conclusion is that if $\xa$ is a Killing field,
then perturbations about the background spacetime obey an integral constraint.
The integration volume is now a Lorentzian slice, so its
boundary includes  initial and final spacelike surfaces, plus the boundaries at
infinity and along a horizon if present. In general  there
are fluxes through the initial and final slices. However, if
the spacetime and the perturbations are static, the general statement
can be reduced to the constraint \introlaw\ on an  asymptotically
flat spatial slice . 

The difference between the constraints derived by slicing the spacetime with a timelike normal
field, or with a spacelike normal, is  similiar to the
difference between Gauss' Law and Ampere's Law in Maxwell theory. Start with
Maxwell's  equations in covariant form, $\nabla _a F^{ab}=J^b $.
Then the difference between the derivation of
the two laws is just the choice of splitting spacetime into
space plus time, or into space plus a smaller spacetime, that is, taking the free index ``$b$' to
be a time direction or a space direction respectively. 
 Gauss's Law is a true
constraint, whereas Ampere's law holds
 only if the electric field is time
independent. Even so, Ampere's law  is still a useful relation. 
 In Gauss's law, the
source is the time component of the charged current; in Ampere's law, the spatial components
contribute. Similiarly, in the gravitational construction
the conserved charges $ Q^{tab}$ contribute to the mass, while
the $ Q^{xab}$ contribute to the tension. In general this will include purely spatial components of
the charged current. Having outlined the idea, we now proceed with the details of the calculation.

\subsec{The d\ +\ 1 Split}  

 Let M denote a
$d+1$-dimensional manifold  with metric $g_{ab}$ of signature $(- \,+ \,+ \cdots +)$. Let
$\nabla_a$ be the derivative operator compatable with $g_{ab}$,
$\nabla_a g_{bc}=0$. 
Let $\Sigma_w$ be a family of $d$ dimensional submanifolds with constant coordinate $w$,
possibly defined just in some subset of M. Let $ n =n\cdot n N  dw$ be the unit normal to $\Sigma
_w$, where
 $n\cdot n \equiv n^a n_a=\pm 1$ if $\na$ is spacelike or timelike respectively. We want to rewrite
the Einstein equation in terms of $d$ dimensional quantities. Here we keep track of various minus
signs that distinguish between a spacelike and a timelike foliation. The details are omitted; the
steps for the usual $3+1$ decomposition can be found in standard texts \mtw\ \waldbook  .

The metric $g_{ab}$  induces a metric $s_{ab}$ on  $\Sigma_w$ 
\eqn\metricsplit{
g_{ab}=s_{ab} +(n\cdot n) n_a n_b \  , \   \    \na s^{ab}=0}
and let $D_a s_{cd}=0$. The lapse $N$ and shift $N^a$ are defined by the decomposition
 $${\partial \over \partial w} =W^a = Nn^a + N^a \  ,\  \na N^a =0 $$

We will take the Einstein Lagrangian to be $L_G =+R$.
 The momentum conjugate to $s_{ab}$ is
\eqn\gravmomen{
\pi^{ab} = (n\cdot n) \sqrt{|s|} ( K s^{ab} - K^{ab} )} 
where $K_{ab}=s_a^c \nabla_c n_b$ is the extrinsic curvature of $\Sigma_w$.
The projection of a tensor
onto the submanifold $\Sigma _w$ will be denoted by
 $\tilde{A}^b \equiv s^b _c A^c$. When no ambiguity
arises we will drop the "tilde", for example,  from their definitions, the shift, extrinsic curvature,
and momenta fields are tangent to $\Sigma$.

 By considering a spacetime decomposition as introduced above
one finds that the components $G_{ab} n^b$ of the Einstein tensor 
only involve first derivatives in the $w$
coordinate. When $W^a$ is timelike then the components $G_{ab}n^b = 8\pi T_{ab}n^b$ are known
as the Einstein constraint equations. Here we will refer to these equations as constraints
whether $W^a$ is timelike or spacelike.
Explicitly,  
\eqn\einstcons{\eqalign{
(n\cdot n) H_G &= 16\pi T_{ab} n^a n^b  \cr 
(n\cdot n) {H_G}^b &=16\pi s^{bc} T_{cd} n^d  \cr }} 
where
\eqn\defcons{\eqalign{
(n\cdot n) H_G &= 2G_{ab}n^a n^b
=  - (n\cdot n)R^{(d)}  + {1 \over |s|}  ({\pi ^2 \over d-1 } - \pi^{ab} \pi_{ab} ) \cr
(n\cdot n){H_G}^b & =  {H_{G}}^b = 2  s^{bc}G_{cd} n^d= -2(n\cdot n) D_a (|s|^{-{1 \over 2}} \pi^{ab} ) \cr }}

The Hamiltonian for Einstein gravity is pure constraint. Variation of
the Hamiltonian density
\eqn\einstham{ {\cal H}_G =\sqrt{|s|} (NH_G +N_b H^b_G ) }
yeilds the vacuum Einstein equations. In the Hamiltonian 
variation, $''q"$ is $s_{ab}$ and $''p"$ is $\pi ^{ab}$.
$N, N^b$ are Lagrange multipliers. 

In this paper, we consider General Relativity coupled to matter
fields whose stress energy is derived from a Lagrangian $L_M$. We will assume
 that the matter $L_M$ can be transformed to Hamiltonian form.
 The particular examples we work out have
 total Lagrangian  $L= R-F^2$ with $F =dA$. 
  The matter Hamiltonian $H_M$ may contain additional constraint
equations ${\cal C}=0$, for example, Gauss' Law.  
The total Hamiltonian for the coupled system is 
\eqn\totham{ (N, N^a , \alpha )\cdot{\cal H}_{tot} (g,\pi , A, p )
 =\sqrt{|s|} [ N (H_G +H_M ) +N_b (H^b_G +H_M ) ] +\alpha {\cal C}}
where  $\alpha$ is a Lagrange multiplier for the matter constraints.
Equations \einstcons\ and ${\cal C}=0$ imply that
on solutions to the field equations,
\eqn\constwo{  (N, N^a , \alpha )\cdot {\cal H}_{tot} (g,\pi , A, p )=0}

\subsec{Derivation of Constraints}

The derivation of constraints on perturbations  \jt\  \kt\  which we use
is based upon the fact that the gravitational Hamiltonian density is a sum of constraints, \constwo\ . This previous work includes an analysis 
of the Einstein equation with a fluid stress energy, applied to cosmological
spacetimes. The fluid sources are described by a stress-energy tensor, and not derived from a
Lagrangian; typically the fluid is described by a density, pressure, and velocity, rather than some more
fundamental set of fields. In this paper we apply the same constraint vector construction for
Einstein gravity coupled to  matter fields which are described by a Lagrangian. 
The derivation is similiar to calculations
of linearization instability \arms\  \amm\  \mon\  . 
The linearization instability work was done for compact manifolds without boundary. Here we work on
non-compact manifolds, and the boundary terms are  of particular interest, since they
are the mass, horizon area, and tension variations.

Gauss' Law  $\nabla_a F^{abct} = J^{bct}$ is a nice, linear constraint, which can be rewritten in 
integral form. The following construction extracts a similiar Gauss' Law type statement from the
{\it linearized} Einstein constraints  , in the case that the background spacetime has
a Killing field. 
Let $(\bar{s}_{ab} \ ,  \bar{\pi}^{ab}, \bar{A}_{abc},\bar{p}^{abc}) $ be a solution to the field equations,
where $p^{abc}=8\sqrt{|s|} NF^{abcw}$ is the momentum conjugate to $A_{abc}$ (see Appendix ).
Consider j perturbations about this solution,
\eqn\defperturb{\eqalign{
q's \  :\  s_{ab} &= \bar{s}_{ab} +\epsilon h_{ab} \ ,\    A_{abc}= \bar{A}_{abc}+\epsilon \delta A_{abc} \cr
p's \  :\  \pi^{ab}& = \bar{\pi}^{ab} + \epsilon\delta \pi ^{ab} ,\  p^{abc}= \bar{p}^{abc} +\epsilon \delta p^{abc} \cr
}} 
where $\epsilon\ll 1$ .The result is derived by finding  a judicious linear combination of the linearized  constraints.
The idea is very simple, and to display this fact, we streamline the notation:
Define the variation $\delta$ by $\delta f =
 {\delta f \over \delta s_{ab}}\cdot h_{ab} + {\delta f \over \delta \pi^{ab}}\cdot \dpi^{ab} +
{\delta f\over \delta A_{bcd}}\cdot \delta A_{bcd} + {\delta f\over \delta p^{bcd}}\cdot \delta p^{bcd}$.

Let $F, \beta ^a$ be an arbitrary function and vector on $\Sigma _w$, and
consider the linear combination of constraints
 $ (F,\beta ^a , A_{wbc} )\cdot {\cal H}_{tot}(g,\pi , A, p )
=0$. The perturbative fields, which we have divided above into $\delta q's$ and $\delta p's$, are 
solutions to the linearized constraints
\eqn\linconst{(F, \beta ^a , A_{wbc} )\cdot \delta\htot  \cdot (\delta q ,\delta p ) =0 .}
Rewrite \linconst\  in terms of the adjoint operator and a total derivative,
\eqn\adjoint{  ( \delta q ,\delta p)\cdot \delta \htot ^*\cdot (F, \beta ^a , A_{wbc})  +D_a B^a = 0 }
where $B^a$ is a function of the $\delta q ,\delta p$ , the Lagrange multipliers, and of course the
background spacetime.

Therefore if $F,\beta ^a$ are solutions to the differential equation
  $\delta \htot ^*\cdot (F, \beta ^a, A_{wbc})=0$, then 
 any perturbation about the background spacetime must satisfy the Gauss's Law type constraint 
\eqn\difficvlaw{D_a B^a =0 \   .}
When do solutions for $F, \beta ^a$ exist? Hamiltonís equations are
$(\dot{s} ,-\dot{\pi }  ,\dot{ A} ,-\dot{p} ) = \htot ^*\cdot (F, \beta ^a , A_{wbc})$, where
$\dot{f}$ is the lie derivative of $f$ along $V^a$,
\eqn\icv{V^a=F n^a +\beta ^a}
So the requirement that \difficvlaw\   hold for all perturbations about the background is that the lie
derivative along $V^a$ of all the $q's$ and $p's$ is zero, that is, $V^a$ must be a Killing vector.

The boundary term vector is the sum of a gravitational piece and a contribution from
the matter fields, $B^a = B^a _G +B^a _M$, where
\eqn\gravbt{
B^a _G = F(D^a h - D_b h^{ab}) - h D^a F + h^{ab} D_b F
+ {\beta ^b \over \sqrt{|s|} } (\pi^{cd} h_{cd} s^a_b - 2 \pi^{ac} h_{bc} -2\dpi^a_b ) }
\eqn\mattbt{\eqalign{
B^a _M &=  - { 1\over \sqrt{|s|}} A_w \delta p^a + 
4F\tilde{F}^{ab} \delta \tilde{A}_b +{ 2\over \sqrt{|s|}} 
\beta^{[a}p^{b]} \delta \tilde{A}_b \  ,\  1-form \cr 
B^a _M &= {- 3\over \sqrt{|s|} } \tilde{A}_{bcw} \delta p^{abc} + 8 F \tilde{F}^{abcd} \delta \tilde{A}_{bcd}
 + {2\over \sqrt{|s|} }\beta ^{[a} p^{bcd]} \delta \tilde{A}_{abc} \  ,\   3-form\cr  }}
 Here $h = h_{ab} s^{ab}$ and all indices are raised and lowered using $s_{ab}$. See 
 Appendix A for
some details of the Hamiltonian decomposition of $F^2$.

It is simple to include perturbative sources $\delta T_{ab}^{(s)}$, i.e, stress energy which
does not come from a Lagrangian. Equation \linconst\  is replaced by
\eqn\withsource{ (F, \beta ^a , A_{wbc}
)\cdot \delta\htot  \cdot (\delta q ,\delta p ) = \dsource }
where $\dsource \equiv 16\pi (n\cdot n) \delta T^a _b V^b n_a - 4!A_{bcw}\delta J^{bcw} $, where 
 the charged current is defined by
\eqn\divf{\nabla _a F^{abcd}=J^{bcd}}

It follows that 
when $V^a=F n^a +\beta ^a$ is a Killing vector, then $D_a B^a =\dsource$. In integral form, this is
\eqn\icvint{
   \int _{\Sigma } \sqrt{|s|} \dsource  =\int _{\partial \Sigma} da_c B^c }
\icvint\  is the constraint on  perturbations about a spacetime with a Killing field,  when the background
matter fields come from a Lagrangian. Why is this last condition needed? Consider the Einstein equation
with a fluid stress-energy, as for cosmological spacetimes. In a $d+1$
split, the form of the Einstein constraints \einstcons\  still holds, but Hamiltonís equation for $\dot p$ becomes ${\delta\htot ^* \over
\delta \pi ^{ab} }=-\dot{\pi}^{ab} +8\pi NT^{ab}_{fl}$. One finds that the vector $V^a$ must satisfy a set
of equations different from Killings equation, for a constraint of the form \icvint\  to hold on
perturbations \jt\  . $V^a$ was referred to as an integral constraint vector in that work, and
the Freidman-Robertson-Walker spacetimes turn out to have constraint vectors. Tod \tod\  has shown that
there exists a maximal set of solutions for $V^a$ when $\Sigma$ can be locally embedded in a space of constant
curvature, and that in $3+1$ dimensions this is related to the existence of three-surface twistors.

\newsec{First Law}

In this section we use \icvint\  to give a simple proof of the First Law for black branes.

Let $M$ be a black brane spacetime with a bifurcate Killing horizon, and 
let $V $ be an asymptotically flat spacelike slice with timelike normal $n^a$.
 Assume that the
directions  tangent to the brane are compact. For the non-compact case one must specify
additional
 boundary conditions at infinity along the brane. A simple example is the two-parameter
  family of $M2$-brane spacetimes \cvetic\
\eqn\mtwo{\eqalign{ds^2 &= f^{-2/3} (-H dt^2 +dx^2 +dy^2 ) + f^{1/3} ({dr^2\over H}
+\delta_{ij}dx^i dx^j ) \cr
A_{txy}&= -{\tilde q \over r^6 +q} +{\tilde q \over q} \cr }}
where $f(r)=1+ q/r^6  $, $H(r)=1-q_o /r^6  $. The parameters are related by
 $q=q_o sinh^2 \delta ,\  
\tilde q = q_o sinh\delta cosh\delta $. 
$i,j=1,...,7$ are the transverse coordinates. The extremal, supersymmetric case occurs when $q_o =0$.

 First consider a slice $V_0$ which intersects the horizon $\hor$ on the
bifurcation  surface. For example, the bifurcation surface of
 the $M2$-brane \mtwo\   has topology $S^7$ cross
$T^2$. Assume that the spacetime has a stationary Killing field
 $\xia =Fn^a +\beta ^a$ and substitute $\xia = V^a$
in the constraint relation \icvint\ .  Using the asymptotically flat boundary conditions, 
the boundary terms at infinity simplify, and have standard interpetations as the change in the mass
and the momentum,
 see e.g., \teit\ . The gravitational boundary term \gravbt\   becomes
\eqn\infgravbt{\eqalign{
 - 16 \pi \delta M &= \int_{\partial V_{\infty}} da_c [ F (D^c h - D_b h^{cb} ) -hD^cF +h^c_b D^b F ] \cr
16 \pi \Omega\delta {\cal J_G} &=-\Omega \int_{ \partial V_{\infty}} da_c{2 \phi^b \dpi^c_b \over
\sqrt{|s|} } \cr}}
The terms $-hD^cF +h^c_b D^b F$ do not contribute to the boundary around an asymptotically flat black hole, but
 they may contribute in a black brane spacetime which includes compact dimensions.
 For example, if the spacetime has topology of
a compact manifold $K$ cross $R^n$, then $F$ may depend on the coordinates $z^i$ on $K$.
From the lower $N$-dimensional point of view, the $d+1$ dimensional $\delta M$ would 
break up into an $N$-dimensional $\delta M^{(N)}$ plus contributions from the variations in
moduli fields.

Suppose that the background spacetime has no magnetic charge,
$\int _{\partial V_\infty } \tilde F _{abcd} =0$. Then the integrand
$da_d F\tilde{F}^{dabc} \delta \tilde{A}_{abc}$ vanishes at infinity.
So the contibution to the boundary term at
infinity  from the gauge field is
 \eqn\threeformbt{\eqalign{
<A_{tab}\delta Q^{tab} > &= 3\int_{\partial V_{\infty}} da_c A_{tab}{\delta p^{cab} \over \sqrt{|s|}}
\cr
 16 \pi\delta {\cal J}_M &=  \int_{\partial V _{\infty}}da_c 
 {2\over \sqrt{|s|} }\phi ^{[c} p^{abe]} \delta \tilde{A}_{abe} \cr }}
Here $Q^{tab}$ are the electric charges, $\delta Q^{tab}= 3\int_{(\partial \Sigma)_{\infty}}
 da_c {\delta p^{cab} \over \sqrt{|s|}}
(=4!\int _{\Sigma} \delta J^{tab})$. The last equality just reminds the reader of the relation
between the electic field and charged sources, if the electric field was generated by  smooth sources.

 ${\cal J}_M$ is the gauge  contribution to the angular
momentum. If the gauge potential is constant on the boundary at infinity, then $<A_{tab} \delta Q^{tab} >= A_{tab}\delta
Q^{tab}$; otherwise the charge term in the First Law is an average
over the internal space.

On the bifurcation surface of the horizon $\xia$ vanishes, and the gavitational boundary term is
\eqn\horgravbt{
     \int_{\partial V_0} da_c ( - h D^c F + h^{cb} D_b F ) = 2 \kappa \delta A}
where $\kappa$ is the surface gravity  and $A$ is the area of $\partial V_0$

When $\hor$ is a Killing horizon then it can be shown that for the one-form
gauge potential, $A_a\xia =constant$ on all of $\hor$, see e.g.,\carter\ \sorkin\ . We have not been
able to generalize this result to higher-form potentials. However, when the horizon
has a bifurcation surface then since $\xia=0$, also $A_{abc}\xia =0$  there, as was used in
\horowitz\  for black strings.
 Then all gauge terms vanish  on the horizon, and \icvint\  gives the First Law, 
\eqn\firstlaw{
\delta M =  {\kappa \delta A \over 8 \pi}+ \Omega ( \delta{\cal J}_G  +\delta{\cal J}_M ) 
-{3\over 2\pi} <A_{bct} \delta Q^{bct}>  + \int _{V_o} {(- \dsource )\over 16 \pi } }
Here
 ${-\dsource \over 16 \pi} = \xi ^a n^b \delta T_{ab}
 + {3\over 2\pi} A_{bct} \delta J^{bct} $ . \firstlaw\  assumes that there is
no magnetic charge and that the spatial slice $V_o$ intersects $\hor$ at the bifurcation
surface. The perturbations are arbitrary, i.e., not necessarily stationary.

In the magnetic case, there is a boundary contribution from $ \tilde{F}^{abcd}\delta \tilde{A}_{bcd} $ . This
is because the gauge potential is defined in patches, so when using Stokes theorem to change the volume term to a
boundary term, there are contributions from the boundaries of different patches.
 In the spherically symmetric case, $ds^2 =m_{ab}dx^a dx^b +r^2 d\Omega _4 ^2$, and $F=Q_B \omega _4$, where $\omega _4$
is the volume form on  the unit  four-sphere, this boundary term gives a contribution proportional to $\delta Q_B$:
the integral factors into $Q_B V_I\int dr r^{-4} \sqrt{-m} \int 
 d\alpha d\beta d\gamma 
(\delta A^N _{ \alpha \beta\gamma} - \delta A^S _{ \alpha \beta \gamma} )$, where $\alpha ,\beta ,\gamma$ are 
coordinates on the $S^3$ boundary of the $N$ and $S$ patches, and $V_I$ is the volume of the remaining 
internal space. The integral over the $S^3$ is $\delta Q_B$, but it would seem that one needs to 
know the actual metric funtions to evaluate the remaining integral $dr$, even in this simple case. It would be interesting
to know if the magnetic contribution could be evaluated in general.

\subsec{Extension to General Slices}
The First Law \firstlaw\  was proved for arbitrary perturbations, but restricted to the case when the spatial slice intersects
$\hor$ on the bifurcation surface. 
It is clearly more difficult to evaluate the boundary term on a general cross section of the horizon,
since $\xia$ does not vanish, and the momentum of the spatial slice $\pi ^{ab}$ is also nonzero. However,
we can use the divergence free property of the ICV boundary term as follows.
 Let the family of spacelike slices
$\{ V_w \}$  be asymptoticallly flat, and let $V_0$ be a slice which intersects $\hor $ at the
bifurcation surface.
When there are no fluid sources, \icvint\   becomes
$\int _{\partial V_w} da_c  B^c  =0 $. 
The boundary term vector $B^a$ depends on the geometry of the spatial slice $V_w$ in the
background,  which we will indicate by
"$\epsilon $" (see \defperturb\  ).
 First consider the vacuum case. We have shown that for  any $w$ the contibution to
\icvint\
 from the boundary at infinity is $ -16\pi \delta M +\Omega \delta J$.
 On $V_0$, the boundary term at $\hor$ gives $2\kappa \delta A_0$. Since all the $V_w$ share
the  same boundary at infinity, \icvint\  implies that  the horizon boundary terms have
the same value for all $w$,
\eqn\morefirst{  I (V_w ; \epsilon ) \equiv \int_{\hor _w} B^c da_c = 16\pi \delta M
=2\kappa\delta A_0 }

Sorkin and \sorkin\  have shown that even with time dependent perturbations, the variation in the expansion $\theta$
on the horizon is zero through linear order. This implies that $\delta A _0 =\delta A_w$, ie,
that $\delta A$ is the same on each slice. Therefore \morefirst\ 
implies that $ I (V_T ; \lambda )=\kappa \delta A_T$, since $\kappa$ is constant over the
horizon.

 For Einstein-Maxwell, there is an additional contribution to 
the common boundary term at infinity of $  -4A_t\delta
Q$. What about the gauge term contribution at the horizon? For a one-form potential,
 \carter\  ,\sorkin\  have proven 
that $ A_a \xia $ vanishes on a killing horizon. In appendix B we show that $A_{abc}\xia =0$
on a bifurcate killing horizon; it would be interesting to know if this was true with
just the assumption of a Killing horizon. Therefore, the total boundary term on the horizon
$V_0$ is still $2\kappa \delta A$.
As in the vacuum case, the inner boundary contribution is
fixed at $2\kappa \delta A$. 

\newsec{Constraint on Tension Variations}

Let the background spacetime have a spatial translation
Killing field $\xa$. We want to work out the implications of the general constraint relation 
\icvint\  when the Killing field in the construction is taken to be
$\xa ={\partial\over \partial x}$ and the $d$-dimensional slices $\Sigma$ are surfaces of constant $x$.
The unit normal $\xhat _a = L\nabla _a x $ is spacelike, and $s_{ab}$ is Lorentzian. (For this
decomposition we will write the normal as $\xhat _a$ to distinguish from the
timelike normal $n_a$ used in the last section.) So
\eqn\xsplitmetric{ g_{ab}  = s_{ab} + \xhat _a \xhat _b \  , \qquad \xhat\cdot \xhat =+1 }

As in Section 3, we will assume that the directions tangent to the brane are compact.
 Therefore, we take the spacetime to have topology of $R^N  \times K$, where $K$ is a compact
manifold.  For example, for an $M2$-brane,
$K$ could be an $S^1$ and $N=10$, or $K$ could be a 2-torus cross a four (real) dimensional Kahler
manifold, with $N=5$. 
We also assume that the spacetime is transverse asymptotically flat, $i.e.$,
is asymptotically flat in the noncompact dimensions. The boundary of $\Sigma$ at infinity
is the product of a time interval and the $(d-2)$-dimensional boundary of 
a $(d-1)$-dimensional spatial volume $V_x$, $\partial\Sigma ^\infty = \Delta t \times \partial V_x
^\infty $. The notation $V_x$ indicates  that $x=constant$ in this volume, since
it is a subset of $\Sigma$.
 The rate of fall off of metric perturbations is  $h_{ab}\rightarrow
O(r^{-N+3} )$ as $r\rightarrow \infty$. 

We define the variation in the tension of the spacetime $\delta\mu$ to be to be minus
 the gravitational boundary term
\gravbt\   evaluated at infinity, divided by $\Delta t$. With $X^a = F\xhat ^a +\beta ^a$,
\eqn\defdmu{\eqalign{ 16\pi \delta \mu \equiv -{1\over \Delta t} \int _{\partial\Sigma _\infty} dt 
& da_a [ F(D^a h - D_b h^{ab}) - h D^a F + h^{ab} D_b F\cr
& + {\beta ^b \over \sqrt{|s|} } (\pi^{cd} h_{cd} s^a_b - 2 \pi^{ac} h_{bc} -2\dpi^a_b ) ]\cr}}
The transverse asymptotically flat boundary conditions
imply that the integrand at infinity is independent of time, so the integral over $t$ factors
out, and $\int dt / \Delta t =1$.

\subsec{Perturbations off Minkowski Spacetime}
 We start by considering the integral constraint 
 \icvint\ when the background is flat $(d+1 -p)$-dimensional spacetime,
 cross  a $p$-dimensonal torus. This will give
a physical interpetation of the gravitational boundary term in the weakly gravitating limit.
 Let $\delta T^{(s)b}_a $ be perturbative sources which are localized in the transverse directions. 
 The spacetime volume in \icvint\  is $\Sigma =\Delta t \times R^{d-p}\times I^{p-1}
=\Delta t \times V_x$.  The integration
over the time interval $\Delta t$ cancels. Then
\eqn\dpress{\eqalign{
16\pi \delta\mu & \equiv - \int_{ \partial V ^{\infty}_x} da_c 
  (D^c h - D_b h^{cb}) = 16 \pi \int_{V_x} dV (- \delta p_x ) \cr
& where \qquad  \delta p_x  =\delta T^{(s)b}_a  \xhat ^a \xhat _b \cr } }

The volume integral is minus the $x$-component of the pressure of the source,
$i.e.$, it is the $x$-component of the tension.
Therefore it is reasonable to interpet the boundary term at infinity as  the
spacetime tension.  More precisely, $\delta \mu $ is the integral
of the tension.  This is (minus) the force per unit
$(d-1)$-dimensional area, times the area, $i.e.$, $\delta \mu$ is the
$x$-momentum flux.
 The term tension rather than
pressure is being used here since test $p$-branes governed by the area action
 have positive tension. We emphasize that the tension is in a direction
tangent  to the source, and is not a radial pressure as in a star.

\subsec{ Background Magnetic fields}
One contribution to $\delta M$ is the variation of the electric charge. This occurs because $\delta M$
is associated with the time translation symmetry of the background, and the Lagrange
multiplier of $\delta Q$ is the time component of the gauge potential.
 When the ICV construction is done using a spatial translation
symmetry of the background, then there is a contribution from the spatial
component of the  field strength instead. Consider the case when a test brane moves in a static
background magnetic field, which was studied for a test two-brane in flat spacetime in \taylor\  .
Assume that there is no background electric charge.
 With out gravity, this is the situation for Ampere's Law. For example, suppose that there is
a background gauge potential $A_{xyz} (x^i )$, and
 a test two-brane
that lies in the $x-y$ plane and moves in the $z$-direction. The brane generates a current $\delta
J^{xyz}$, which in turn induces a change in the magnetic field $\delta F^{cxyz}$ satisfying $\int
_{\partial D} da_c \delta F^{cxyz} = \int _D dv \delta J^{xyz}$. 
Including the gravitational field, then a 
perturbative  current $\delta J^{abc}$ generates a perturbation to the gauge field, and this causes a
variation in the gravitational fields. Equation \icvint\  becomes
\eqn\amp{ 16\pi \delta\mu  = -4!\int _{\partial V_x} da_b N A_{cdx} \delta F^{bcdx}
 +\int _{V_x} [ -16\pi \xa \delta T^{(s)b}_a \xhat _b +4!A_{bcx }\delta J^{bcx} ] }
In the example above, there is a gauge field boundary term from a nonzero $A_{xyz}\delta F^{xyz i}$,
rather than from a baackground electric charge.

\subsec{Perturbations off Black Branes}
Now we are ready to
 study perturbations of the spacetime tension in a black brane spacetime. Assume that
the spacetime has  a static Killing field $\xia$, as well as the spatial 
translation killing vector $\xa$, so that the horizon is a
Killing horizon. For simplicity we will specialize to the case where the slicing can be
chosen such that $\xa = L\xhat ^a$ , $g_{xx} =\xa
X_a =L^2$. Let $L_H$ and $L_\infty$ equal length of
the $S^1$, to which $X^a$ is tangent, at the horizon and at infinity respectively. Let $A$ equal to the
area of the cross section where 
 the spatial slice $V$ intersects the
horizon, and let  $A_x$ be the area of the intersection of the subsurface $V_x$ with $\hor$, then  
\eqn\areax{ A=L_H A_x \  .}
 We will also assume that $\xa $ is tangent to the horizon, $\xa \xi _a =0$ on $\hor$.

First consider the contribution to \icvint\   from the boundary on the horizon.
In a neighborhood of the horizon it is useful to write the metric in terms of 
null coordinates. Let $k^a ={\partial\over \partial \lambda}$ and $q^a ={\partial\over
\partial U}$ be null and geodesic, where $k^a$ a geodesic generator of the horizon 
and $\lambda$ is an affine parameter. On $\hor$, 
 $\xia =\kappa \lambda k^a$ . Using these basis vectors the metric is
\eqn\metrichor{
s_{ab} =\gamma _{ab} -\xi _a q_b -\xi _b q_a }
where $\gamma _a^{\ b}\xia =0$, $\gamma _a^b q^a =0$ and $k_a k^a = q_a q^a =0$.  We normalize $q^a$ by
$\xia q_a =-1$ on  $\hor$. Our assumptions on $\xa$ are
\eqn\assume{\xa =L \xhat ^a \  , \xa =\gamma ^a _b X^b \qquad on\  \hor  .}

The gravitational boundary term \gravbt\  depends on derivatives of $L$. On the horizon,
expand the gradient of the
norm $X\cdot X$ in the above basis. Since $\xia ,\xa$ are Killing vectors and
$\xia$  is null, the expansion is of the form
\eqn\xxexpand{ \nabla^a(X_b X^b)=2L\nabla ^a L = -2 \nu \xi^a  +\gamma ^a_{\ b} C^b } 

To evaluate the boundary term on the horizon, we will see that one only needs the form of the expansion
\xxexpand\  and not the values of $\nu$ and $C^a$. The area element on $\hor$ is $da_b = -k_b \rootgamma
d\lambda dy$, where $dy$ indicates integration over all the spatial coordinates except $x$. Substituting
\xxexpand\   into \gravbt\  ,the integrand is
\eqn\horintone{\eqalign{
 -  k_a B^a & = -{1\over \kappa\lambda} \xi_a B^a   \cr
 & = -{L\over \kappa\lambda } [ (\xi_a D_b h^{ab} - \xi_a D^a h)
    + {1 \over L}  (\mu \xi_a  \xi_b h^{ab} + \xi_a h^{ab} C_b)  ] \cr }}

To evaluate this, we make the gauge choices $\delta \xia =0$ and $\delta q^a=0 $ . These are
consistent conditions when the spacetime is static \foot{ When the background
spacetime is stationary, the Killing field which generates the horizon is a linear
combination of the time translation and rotational Killing vectors
 $\xi ^a = t^a +\Omega \phi ^a$.
Then a superior choice of gauge conditions for evaluating the boundary term 
are $\delta t^a =0, \delta \phi ^a =0$ 
 \bch\carter .}.  Since the hypersurface is fixed, we also have that  $\delta n_a  =0$. 
The inner product conditions which define \metrichor\  hold in the background and the perturbed spacetime.
Using these conditions, one finds that on $\hor$, $\xia h_{ab}=0$. So the
integrand on the horizon  reduces to
\eqn\horinttwo{
 -   k_a B^a   = -{L\over \kappa\lambda } [ ( D_a (\xi _b h^{ab} ) -
\xi_a D^a h)]  }
where we have used the fact that $\xi$ is a killing vector.

The spacetime volume $\Sigma _x$  in \icvint\  is a slice of constant $x$. $\partial \Sigma _x$ is the sum of
initial and final timelike slices, a null boundary over part of the horizon,
and a timelike boundary at infinity. At infinity the spacetime approaches the
direct product of
 asymptotically flat Minkowski spacetime cross a (static) compact
space K. The integration at infinity reduces to an integral over the spatial boundary times a time
interval $\Delta t$, where $t$ is the asymptotic killing time parameter. In general, the initial and
final slices will contribute to relation \icvint\   , which therefore is a time-dependent flux
balance statement. There does not appear to be  much one can say about this general
 case. However, if the background spacetime $and$ the perturbations are static,
 then we can extract an interesting statement, which we now turn to.

\subsec{Static Perturbations: Reduction to a Spacelike Slice}

In this section, we assume that the perturbations are static, as well as the background. This allows
reduction of  the integral constraint on $\delta\mu$ to a statement on a spacelike slice $V_x$.
The perturbations need not be translationally invariant.
We choose $\Sigma $ as follows. Pick an
asymptotically flat spacelike slice, which can intersect the horizon at an arbitrary cross
section. Choose  $ (\partial \Sigma ) _i$, the initial spacelike boundary of $\Sigma$, to be a volume 
 contained in
this slice, with inner boundary $R_{in}$ and outer boundary $R_{out}$. We will eventually let
$R_{in}$ go to the horizon and $R_{out}$ to spatial infinity. The static
Killing field  $\xia ={\partial\over \partial t}$ is timelike in this volume. Define the
spacetime volume $\Sigma$ by lie dragging $ (\partial \Sigma ) _i$ an amount $\Delta t$
along the flow of $\xia$ to $( \partial
 \Sigma )_{f}$. Then for static perturbations, the integral of the boundary term over
the final time slice is equal in magnitude to the integral
over the initial time slice, and so these terms cancel in the 
integration over $\partial \Sigma$.

The outer boundary of $\Sigma$ includes an integration over an interval $\Delta t$.
 Take the inner boundary to the horizon, which includes integrating over  an interval of
affine parameter $\Delta \lambda$. On $\hor$,
$\xia ={\partial \over \partial s}=\kappa\lambda {\partial\over \partial \lambda}$.
 Flowing along $\xia$
 in a neighborhood of the horizon, one has $d\lambda ={\partial \lambda \over 
\partial s} ds = \kappa \lambda ds$.
Therefore, upon lie dragging the initial spatial surface by $\Delta t
=\Delta s$,
\eqn\intdlambda{
\int _i ^f {d\lambda \over \lambda}=\kappa \Delta t  }

Now we can evaluate the boundary term \horintone\  on the horizon. For static perturbations,
 $\xia D_a h=0$. Using \assume\ Killings equation,  and the gauge condition $h_{ab}\xia =0$ on  $\hor$,
\eqn\sgone{\eqalign{
  D_a (h^{ab} \xi_b) &=  D_c [  h_a^b \xi _b (\gamma ^{ca} -\xi ^c q ^a -\xia q^c) ] \cr
&=   - \xi ^c D_c (h_{ab}\xi ^a  q^b)  -        q^c D_c   (h_{ab}\xi^a \xi^b) \cr}}
The first term vanishes for static perturbations, and  the second term is proportional
to the change in the surface gravity: From the definition of $\kappa$, and the gauge
condition $\delta q^a =0$ , $2\delta \kappa = q^c \nabla _c (h_{ab}\xia \xi ^b ) $.
\sgone\   then implies that 
\eqn\sgtwo{2\delta\kappa =-  D_a (h^{ab} \xi_b) }

Combining \horinttwo\ ,\intdlambda\  ,and \sgtwo\   we can evaluate the boundary term on
the horizon, in the case that the perturbations are static.   We find
\eqn\horbt{
 -  \int \rootgamma d\lambda dy  k_a B^a  = -2\delta\kappa {A\over L_H} \Delta t }
where we have used equation \areax .
Note that the time interval $\Delta t$ will cancel an identical contribution from the boundary
term at infinity, so one is left with a statement on a spacelike slice. This is what one
expects for the static case; the spacetime integral reduces to a statement
which is independent of which time.

The gauge field  boundary integrand is $A_{bcx}\delta p^{bcd}da_d $. If the brane carries
electric charge and is static, then only the tangent components of the gauge potential 
are nonzero. Pecisely, suppose that $A_{bcx} =\Sigma _i f_i \xi _{[b} u_{c]} $. Then
since $da _d \sim \xi _d$ the gauge boundary term vanishes on $\hor$.

Then using \horbt\  , equation \icvint\  becomes
\eqn\dmuone{  \delta\mu =-{A\delta\kappa\over 8\pi  L_H}
 -{3\over 2\pi}<A_{bcx} \delta Q^{bcx}>_x  -\int _{V_x} 
{1\over 16 \pi}\dsource ( \vec{X} ) } 
where
\eqn\xsource{
{1\over 16\pi}\dsource ( \vec{X} ) =  N\delta T^a _b X^b \xhat _a -{3\over 2\pi}A_{bcx }
\delta J^{bcx} \  , }
$L_H$ is defined in equation \areax\ and where $\delta\mu$ is given by

\eqn\defdmu{ 16\pi \delta\mu  =-\int _{ \partial V_x ^{\infty} } da_c B^c_G =
\int _{ \partial V_x ^{\infty} } da_c [ L(D^c h - D_b h^{cb}) - h D^c L + h^{cb} D_b L ] \  . }   
The brackets $<**>_x$ indicate integration over internal dimensions, with $x$ is held
constant, $<A_{xab}\delta Q^{xab} >_x  = 3\int_{\partial V^{\infty}_x} da_c A_{xab}{\delta p^{cab} \over
\sqrt{|s|} }$.
 If the gauge field goes to a constant at infinity, then
$<A_{bcx} \delta Q^{bcx}>_x = A_{bcx} ^\infty  \delta Q^{bcx} $.

Equation \dmuone\  is the main result of this paper. It relates variations in the spacetime tension
to variations in the horizon geometry, the gauge fields, and sources. The derivation of
 \dmuone\   assumes that the spacetime and the perturbations are static, the brane is electric,
 the Killing field $X^a$ is orthoganal to the surfaces of constant $x$, and that $X^a$ is tangent
to the horizon.

The $A\delta\kappa $ term can be replaced with a
 $\kappa \delta A$ term by using a Smarr-Komar type relation,
 as follows. For
a solution to the Einstein equation in $n$ dimensions with static
Killing vector $\xia$, $-{1\over 2}\int _{\partial V} dS_{ab} \nabla ^b \xia =8\pi \int _V dV 
(T^a _b -g ^a _b {T\over n-2}) \xi ^b n_a , \rightarrow 8\pi {n-3\over n-2} \int \rho$ in the
non-relativistic limit. This implies the normalization for the Komar mass, $ 8\pi {n-3\over
n-2} M \equiv
 -{1\over 2}\int _{\partial V} dS_{ab} \nabla ^b \xia $. Applying this Stokes relation
when there is a boundary at a horizon, and with  gauge fields gives
\eqn\smarr{  8\pi {n-3\over n-2} M =\kappa A -6 <A_{bct} Q^{bct}>
+8\pi \int _V L\sqrt{q} (T^a _b -g ^a _b {T\over n-2}) \xi ^b n_a  }
In equation \smarr\ , and in the next few formulae, 
we will need to distinguish between the volume elements on $V$ and $V_x$. These will be
written as $L\sqrt{q}$ and $\sqrt{q}$ respectively.

For simplicity of presentation we will now specialize to the case where there are no gauge fields,
 $A_{abc}=0$. It is straightforward to keep track of all the
gauge field contributions, but the resulting expression is rather lengthy. Instead we will focus on the
gravitational and stress-energy contributions. 
Varying \smarr\  and using the First Law gives 
\eqn\adeltak{ A\delta\kappa =-{\kappa \delta A \over n-2} 
-{8\pi \over (n-2)} \int _V (\delta T^a _b \xi ^b n_a -\xi ^c n_c \delta T) L\sqrt{q}  }
Therefore
\eqn\dmutwo{ \delta\mu  =  {\kappa \delta A \over L_H 8\pi ( n-2)} 
-\int _{V_x} \sqrt{q} \left( N\delta T^a _b X^b \xhat _a +{L\over L_H (n-2)} \int dx
 ( \delta T^a _b \xi ^b n_a -\xi ^c n_c \delta T) \right) }
 Comparing the First Law \firstlaw\  and \dmutwo\  shows that in general
$\delta M$ and $\delta\mu$ are independent physical quantities. 

Actually, we should compare
the mass per unit length to $\delta\mu$ so that the quantities have the same units, since in
defining $\delta\mu$ we divided by the length of the time interval $\Delta t$. Since we are fixing
boundary conditions at infinity, it makes sense to regard the length $L_\infty$ as fixed, and
let $\bar{M}= {M\over L_\infty}$. When $T_{ab}=0$, 
\eqn\puregrav{\delta \bar{M}= {\kappa \delta A \over 8\pi L_\infty }
 \  ,\qquad  \delta \mu = {\kappa\delta A \over ( n-2)8\pi L_H } } 
So under classical processes the tension
increases; under Hawking evaporation, the black brane unstresses.
 
\subsec{Horizon Geometry}
The surface gravity and area are purely geometrical properties of a Killing horizon.
 The definition of surface gravity, and the fact that it is constant, depend on the fact
the horizon is generated by a Killing field. Similiarly,
 one can ask if the existence of the spacelike Killing field $\xa$ implies any
geometrical properties of the horizon?
 Let us recall how the notion of the surface gravity arises.
 On the horizon,  $\xi \cdot
\xi  =0$. The gradient  of the norm defines the surface gravity, $\nabla^a(\xi \cdot
\xi)= -2 \kappa \xi^a $. The gradient does not have terms of the form 
$\gamma ^a _b W^b$ since the vorticity of $\xia $, $\nabla _{[b }\xi _{c]}$, vanishes when
projected onto the horizon. These properties imply that that $\kappa$ is a constant
on $\hor$ \bch\  when $R_{ab}\xia W^b =0$ for all $W^a$ which are tangent to $\hor$.
(So $\kappa$ is constant in vacuum and Einstein-Maxwell black holes, for example .)  $\kappa$ then
appears in the horizon boundary term,  because the boundary integrand involves the derivative of the
norm of the constraint vector.

 We continue to assume equation \assume\ , that $\xa$ is hypersurface orthogonal and tangent to the
horizon.
 Define the vorticity of $\xa$ by $w _{ab} =\nabla _{[a} X_{b]} = \nabla _a X_b$. The vorticity projected onto the
horizon is $\hat w _{ab}= \gamma _a^c \gamma _b ^d w_{cd}$.  \xxexpand\
defines the rate of change of the norm of $\xa$, which is analogous to the definition of
 surface gravity. The last term in \xxexpand\  vanishes if $\hat w _{ab} =0$. While the
vorticity of the null generator $\xia$ necessarily vanishes on $\hor$,
this is certainly not the case for a generic Killing field.
For example, a rotational Killing vector for a spherically symmetric black hole has nonzero  projected
vorticity. However, it is true in cases of interest for a spatial translation --e.g.
in \mtwo\  , the static black M2-brane solutions.

If $\hat w _{ab} =0$, then \xxexpand\ becomes
\eqn\shortnorm{\nabla^a(X\cdot X) |_{\hor}=2L\nabla ^a L = -2 \nu \xi^a \  .}
We will show that with this assumption, then (i)
the norm $X\cdot X$ is constant on $\hor$, (ii) there is a simple formulae for the coefficient $\nu$
in terms of scalar quantitites, and that (iii)  $2\nu \kappa = R_{ab}\xa X^b$ on $\hor$.

 Let ${\cal D} _a$ satisfy ${\cal D} _a \gamma _{bc}=0$.
Then ${\cal D} _c \gamma ^a _b X^b  =\gamma _c ^m \gamma ^a _n \nabla _m X^n + (X\cdot q)
\gamma _c ^m \gamma ^a _n \nabla _m \xi ^n = {\hat w} _c ^a $, since the projected vorticity
of $\xia$ vanishes on $\hor$. Therefore
 ${\cal D} _b
(X\cdot X) =-2 \xa {\hat w}_{ab} = 0$, $i.e.$, the norm is constant on $\hor$.
 In general the norm is a non-zero constant;
for extremal branes wrapped around the $x$-direction, $X\cdot X =\xi \cdot \xi $ which is zero on
$\hor$.

The coefficient $\nu$ in \shortnorm\   which
 describes the rate of change of the norm is given by $\nu =\xa q^b \nabla _b X_a$. This is
ananlagous to $\kappa = \xia q^b \nabla _b \xi _a (=\xia \xi ^b \nabla _a q_b )$. 
The surface gravity can also be computed by the formulae
 $\kappa ^2 = lim{1\over 2} 
\nabla _a \xi _b  \nabla ^a \xi ^b = 
lim  (\xi \cdot \xi )^{-1} ( \xia \nabla _a \xi _b ) (\xi ^c \nabla _c \xi ^b )  
$, where the limit is taken as the ratio approaches the horizon. 
The latter expressions do not contain the basis vector $q^a$.

Using a derivation similiar to the arguements in \carter\ \waldbook\  ,
 we next show that \assume\ and $\hat w _{ab} =0$ imply
\eqn\nusquared{ \nu ^2 = lim{1\over 2} {X\cdot X \over \xi \cdot \xi } \nabla _a X_b  \nabla ^a X^b =
lim {( \xa \nabla _a X_b )( X^c \nabla _c X^b )\over \xi \cdot \xi }\  .}
 Since $\xa$ is a Killing field, 
\eqn\mush{ 3X_{[a} w_{bc]}
X^{[a} w^{bc]}=X\cdot X w_{bc}w^{bc} +{1\over 2} \nabla _c (X\cdot X)\nabla ^c (X\cdot X) .}
Equation \assume\ states 
 that $\xa$ is hypersurface orthogonal, in which
case Froebenius' Theorm states that $X_{[a} w_{bc]} =0$. \mush\  becomes
\eqn\frob{ -(X\cdot X) w_{ab}w^{ab}= {1\over 2}\nabla _c (X\cdot X)\nabla ^c (X\cdot X).}
 Evaluating this on the horizon
with \shortnorm\  implies that $ w_{ab}w^{ab}=0$ .\foot{Note that $w_{ab}w^{ab} =
{\hat w} _{ab}{\hat w} ^{ab} -4\gamma ^{be}(\omega _{bc}q^c )(\omega _{ef} \xi ^f)
-2(\omega _{bc} q^b \xi ^c )^2 $ On $\hor$, $\gamma ^{be} (\omega _{ef} \xi ^f) =0.$
Therefore if $\wab =0$ on  $\hor$, then $\omega _{bc} q^b \xi ^c =0$ and the only
possible nonzero components of the vorticity are $\gamma  ^{ab} q^c \omega _{bc}$. }
 Now, the Froebenius condition implies that the left hand
side of \mush\  is zero, and so is its gradient. The second term on the right hand side approaches
$-2\nu ^2 \xi \cdot \xi$ on the horizon, but its gradient is nonzero. Therefore dividing \mush\  by
$\xi \cdot \xi$ and taking the limit as the horizon is approached gives $0=lim X\cdot X w_{ab}w ^{ab}
(\xi \cdot \xi )^{-1} -2\nu ^2 $, which is the desired result. The second form in \nusquared\  follows
from using \frob\  .

Lastly we show that when ${\hat w}_{ab} =0$ , $2\nu \kappa = R_{ab}\xa X^b$ on $\hor$.
As just noted, on $\hor$ $ w_{ab}w^{ab}=0$. Since $\xa$ is a Killing field, 
 $-R_{ab}\xa X^b =\xa \nabla _c \nabla ^c X_a ={1\over 2} \nabla _c \nabla ^c (X\cdot X)
- w_{ab}w^{ab} = {1 \over 2} \nabla _c \nabla ^c (X\cdot X)=
 -\xia q^b \nabla _a \nabla _b (X\cdot X ) =-q^b
\nabla _b (X\cdot X ) (\xia \xi^c \nabla _a q_b ) =\nu \kappa $, where we have used
the fact that $X\cdot X$ is constant on $\hor$.
So $\nu =0$ in a vacuum spacetime, as long as the surface gravity is non-zero.

\newsec{ Concluding Remarks}
Expressions such as \firstlaw\  for $\delta M$ let us make contact with our Newtonian
intuition, since \firstlaw\  states that one contribution to $\delta M$ is an integral
over $\delta T_{tt}$, though $\delta M$ is defined even when $T_{ab}=0$.
 The idea of the construction in this paper, is to relate the integral of other
components of the stress energy to boundary integrals, possibly with a horizon present.
Foliating the spacetime by $x= constant$ surfaces in the Hamiltonian construction, and
using the Killing vector $\xa ={\partial \over \partial x}$ relates the integral of  $\delta T_{xx}$
to a boundary term. Using this boundary term, $\delta\mu$ is still defined even when $T_{ab}=0$.

There is a tension $\mu _{(i)}$ and a constraint on $\delta \mu _{(i)} $ for each symmetry ${\partial
\over \partial w^i}$. The expressions have different forms for
directions in which a charged brane is wrapped, and directions which are
not wrapped. If $\partial \over \partial x$ is tangent to the gauge potential, then
$\delta Q_{abx }$ enters the constraint  $\delta \mu _{(x)} $, whereas if $\partial \over
\partial w$  is a symmetry direction but is not tangent to the gauge potential, there is no
contribution from the gauge field to $\delta \mu _{(w)} $.
 We see from comparing the first law
\firstlaw\  to \dmutwo\  or  \dmuone\   that in general $\delta M$ and $\delta\mu$ are
independent physical quantities. For example, perturbative sources contribute
differently to the mass and the tension.
s
In order to derive constraints which have the same form as the first law, we assumed
that the spacetime has a spacelike $and$ a timelike Killing vector. 
Analysis of $\delta\mu$ could alternatively be done for
a cosmological spacetime which has spatial isometries, but is not static. 
This might provide just such a contrast as is generally useful for our own instruction,
and our neighbors' entertainment.

\bigbreak\bigskip\bigskip\centerline{{\bf Acknowledgements}}\nobreak 
We would like to thank David Kastor, Rafael Sorkin, Ken Hoffman, 
Herb Bernstein, and Rafael Bousso for many useful
conversations. JT thanks the Institute for Theoretical Physics and the Aspen Center for Physics for their
hospitality.  This work was supported in part by NSF grants PHY98-01875 and Phys-9722614.

\appendix{A}{The Hamiltonian Formulation of $L_M=- F_{abcd} F^{abcd}$ in Curved
Spacetime} Let the total Lagrangian be $L=L_G +L_M = R-F^2$, where
$F_{abcd}= 4 \nabla_{[a} A_{bcd]}$. The metric is split as in \metricsplit\  .
The field coordinate is $\tilde{A}_{abc}= s_a ^{m} s_b ^{n}  s_c- ^{l}  A_{mnl}  $ and
the momentum conjugate to $\tilde{A}_{abc}$ is $p^{abc} ={\delta \sqrt{-g} L_M \over 
\delta \partial_{w} \tilde{A}_{abc} } = -8 N \sqrt{|s|}
\tilde{F}^{wabc} =-8 (n\cdot n) \sqrt{|s|} \tilde{F}^{dabc}n_d $.
Carrying out the standard Legendre transform, the Hamiltonian density is
\eqn\maxham{
\sqrt{|s|} H_{M} = N [{-(n\cdot n) \over 16 \sqrt{|s|}} p_{abc} p^{abc} +
 \sqrt{|s|} \tilde{F}^2 ] + {1 \over 2} N^d
\tilde{F}_{dabc}p^{abc} - 3 \tilde{A}_{bcw} \partial_{a} p^{abc} }
 Hamilton's equations are 
\eqn\hameqs{\eqalign{
\dot{\tilde A}_{abc} & = {\delta \sqrt{|s|} H_{M} \over \delta p^{abc} } =
 {-(n\cdot n )N \over 8 \sqrt{|s|} } p_{abc} + {1 \over 2}N^d \tilde{F}_{dabc} + 3
\partial_{[a} A_{bc]w} \cr  
\dot p^{abc} & = - {\delta \sqrt{|s|} H_M \over \delta
\tilde{A}_{abc} }  = 8 \sqrt{|s|} D_{d} (N \tilde{F}^{dabc} ) +  D_{d} ( N^{[d}p^{abc]} )
\cr          
 0 & = {\delta\sqrt{|s|}H_M \over \delta A_{bcv} } = -3\partial_{a} p^{abc} \cr }}

 In the ICV construction one uses
\eqn\ugh{{\delta \sqrt{|s|} H_M \over \delta s_{ab} } = {(n\cdot n) N \over 16 \sqrt{|s|}}
[{1\over 2} p^{cde}p_{cde}s^{ab}  - 3 p^{acd} {p^b}_{cd} ] + N \sqrt{|s|} 
({1\over 2} \tilde{F}^2  s^{ab} + 4 \tilde{F}^{acde} {\tilde{F}^b}_{cde} ) }
With these ingredients, one finds the boundary term \mattbt\  when performing the variations to
arrive at \adjoint\  .

We also record the corresponding expressions for the one-form gauge potential, which serves to 
completely define the boundary term \mattbt\  . Let $L_M= -F^2$, where
$F_{ab} = 2 \nabla_{[a} A_{b]} $. Then  $p^a= 4(n\cdot n) \sqrt{|s|}
n^c F_{cb} s^{ab} =-4 N \sqrt{|s|} F^{wa}$ is the electromagnetic momentum conjugate to
$\tilde{A}_a$ . The Hamiltonian density is

\eqn\oneham{
\sqrt{|s|} H_M = N [- {(n\cdot n ) \over 8|s| } p_a p^a + F_{ab} F^{ab} ] - N_b p^a {F_a}^b +
-A_w\partial_a p^a }
And Hamilton's equations  are

\eqn\oneeq{\eqalign{ \dot A_a &=
{\delta H_M \over \delta p^a} = - {(n\cdot n) N \over 4 \sqrt{|s|} } p_a - \tilde{F}_{ab} N^b + \partial_a A_w \cr
\dot p^a &= - {\delta \sqrt{|s|}  H _M \over \delta A_a} = \sqrt{|s|}  D_b  4N \tilde{F}^{ba} 
- 2 D_b  p^{[b} N^{a]}   \cr 
0 & = {\delta  H _M \over \delta A_w} = - \partial_i p^i  \cr }}

Using these expressions, one finds the one-form boundary term in \mattbt\  when working out \adjoint\  .
In this case, the charge term in the First Law \firstlaw\  , is replaced by ${1\over 4 \pi} A_t \delta Q$,
where $\delta Q= \int _{\partial V} N da_c F^{ct} =\int _V J^t \  , \nabla _a F^{ab} =J^b$.
The gauge field contribution to the angular momentum is
$  \delta {\cal J}_M = - {1\over 16 \pi}\int_{\partial _V}{da_a \over \sqrt{|s|}} 
  (2 p^{[a} \phi^{b]} \delta \tilde{A}_b  )$

\appendix{B}{Constancy of $p$-form gauge potential  on $\hor$}
Let $\hor$ be a  bifurcate killing horizon, with null generator $\xi$. One can
 choose
a gauge such that $\lxi A =0$, where $A$ is the $p$-form gauge potential.
Let $s$ be a coordinate along an integral curve of $\xi$,
 $\xi ={\partial \over \partial s}$. If $A$
is a one-form, then $\lxi (A\cdot \xi) =0$, which implies that $A\cdot \xi $ is
 independent of $s$ along the curve.
 Then on $\hor$, $ A\cdot \xi =0$, since it is zero on the bifurcation surface.

Next let $A$ be a two-form, and define the one-form $u_a =A_{ab}\xi ^b$. Then
$\lxi u =0$, and $\lxi u\cdot u =0$. So the norm $u\cdot u$ is a constant independent
of $s$ on each integral curve of $\xi$. Since $u\cdot u =0$ on
the bifurcation surface, it follows that everywhere on $\hor$,$u\cdot u=0$.
 But now the inner product is  positive definite, since the metric
 can be written as $g_{ab}=\bar\gamma _{ab} -\xi _a q_b -\xi _b q_a $
 (similiar to \metrichor\  ). So $u \cdot u= (\bar\gamma _{ab} -\xi _a q_b -\xi _b q_a  )
A^a _{c}\xi ^c A^b _{d}\xi ^d =\bar\gamma _{ab} u^a u^b .$ 
Therefore $u_a =0$ on $\hor$. 
The case for a three-form gauge potential proceeds in the same way.

\listrefs

\end